\documentclass[10pt]{article}

\usepackage{arxiv}

\usepackage[utf8]{inputenc} 
\usepackage[T1]{fontenc}    
\usepackage{hyperref}       
\usepackage{url}            
\usepackage{booktabs}       
\usepackage{amsfonts}       
\usepackage{nicefrac}       
\usepackage{microtype}      
\usepackage{lipsum}
\usepackage{graphicx}
\usepackage{amsmath} 
\usepackage{textcomp}

\title{Designing nanophotonic structures using conditional-deep convolutional generative adversarial networks}

\author{
  Sunae So\\
  Department of Mechanical Engineering\\
  Pohang University of Science and Technology (POSTECH)\\
  Pohang 37673, Republic of Korea \\
  \And
  Junsuk Rho\thanks{Corresponding author: jsrho@postech.ac.kr} \\
  Department of Mechanical Engineering, \\
  Department of Chemical Engineering\\
  Pohang University of Science and Technology (POSTECH)\\
  Pohang 37673, Republic of Korea \\
}

\begin{document}
\maketitle

\begin{abstract}
Data-driven design approaches based on deep-learning have been introduced in nanophotonics to reduce time-consuming iterative simulations which have been a major challenge. Here, we report the first use of conditional deep convolutional generative adversarial networks to design nanophotonic antennae that are not constrained to a predefined shape. For given input reflection spectra, the network generates desirable designs in the form of images; this form allows suggestions of new structures that cannot be represented by structural parameters. Simulation results obtained from the generated designs agreed well with the input reflection spectrum. This method opens new avenues towards the development of nanophotonics by providing a fast and convenient approach to design complex nanophotonic structures that have desired optical properties.
\end{abstract}


\section{Introduction}
 Progress in nanophotonics has yielded numerous extraordinary optical properties such as cloaking objects \cite{cai2007optical, schurig2006metamaterial}, imaging beyond the diffraction limit \cite{zhang2008superlenses, betzig1992near}, and negative refractive index \cite{shelby2001experimental, valentine2008three}. In nanophotonics, a sub-wavelength antenna interacts with light, so precisely-designed components can provide useful functionalities. However, the field of nanophotonics lacks a systematic protocol to design desired components \cite{matlack2018designing, kalinin2015big, coulais2016combinatorial}, so the process still relies on a laborious method of optimization. Such conventional design method requires time-consuming iterative simulations. 
 
 Recently, data-driven design approaches have been proposed to overcome this problem. These approaches use artificial neural networks to design nanophotonic structures \cite{liu2018training, malkiel2017deep, kabir2008neural, ma2018deep}. Previous studies first set the shape, such as multilayers \cite{liu2018training} or H-antenna \cite{malkiel2017deep} of structures to be predicted, then trained NNs to provide output structural parameters that achieve desired optical properties. Once the NNs are trained, they provide the corresponding design parameters without additional iterative simulations. Such attempts have greatly reduced the effort and computational costs of designing nanophotonic structures. So far, these approaches have only been applied to conditions in which basic structures are predefined where only structural parameters are predictable. Most recently, a generative adversarial network (GAN) model have been used to inversely design metasurfaces in order to provide arbitrary patterns of the unit cell structure\cite{liu2018generative}.
    
 In this paper, we provide the first use of conditional deep convolutional generative adversarial network (cDCGAN) \cite{radford2015unsupervised} to design nanophotonic structures. cDCGAN is recently developed algorithm to solve the instability problem of GAN, which provides much stable Nash equilibrium solution. The generated designs are presented as images, so they provide essentially arbitrary possible design for desired optical properties which are not limited to specific structures. Our research provides designs of a 64 $\times$ 64 $-$ pixels probability distribution function (PDF) in a domain size of 500 nm $\times$ 500 nm, which allows $2^{64\times 64}$ degrees of freedom of design.
 
 Artificial intelligence has revolutionized the field of computer vision \cite{russell2016artificial, krizhevsky2012imagenet}. A convolutional neural network (CNN) \cite{krizhevsky2012imagenet, lawrence1997face} is among the most widely used techniques, inspired by the natural visual perception mechanism of the human brain. A CNN uses convolution operators to extract features from input data, which are usually images. It greatly increases the efficiency of image recognition, because feature maps extract important features of the images. The development of GAN has yielded major progress in computer vision \cite{goodfellow2014generative}. A GAN consists of a generator network (GN) that generates images, and a discriminator network (DN) that distinguishes generated images from real images. GN is trained to generate authentic images to deceive DN, and DN is trained not to be deceived by GN. The two networks compete with each other in every training step; ultimately, the competition leads to mutual improvement of each network, so that GN can generate higher quality realistic images than when it learns alone. DCGAN combines the idea of a CNN and GAN to provide much stable Nash equilbrium solution \cite{radford2015unsupervised}. cDCGAN replaces fully-connected layers with convolutional layers \cite{radford2015unsupervised} with condition \cite{mirza2014conditional}, which is target reflection spectrum in our case.

\section{Results and Discussions}

\subsection{Deep-learning procedure}
\label{sec:Deep-learning}

    \begin{figure}[h!]
    \centering\includegraphics[width=13cm]{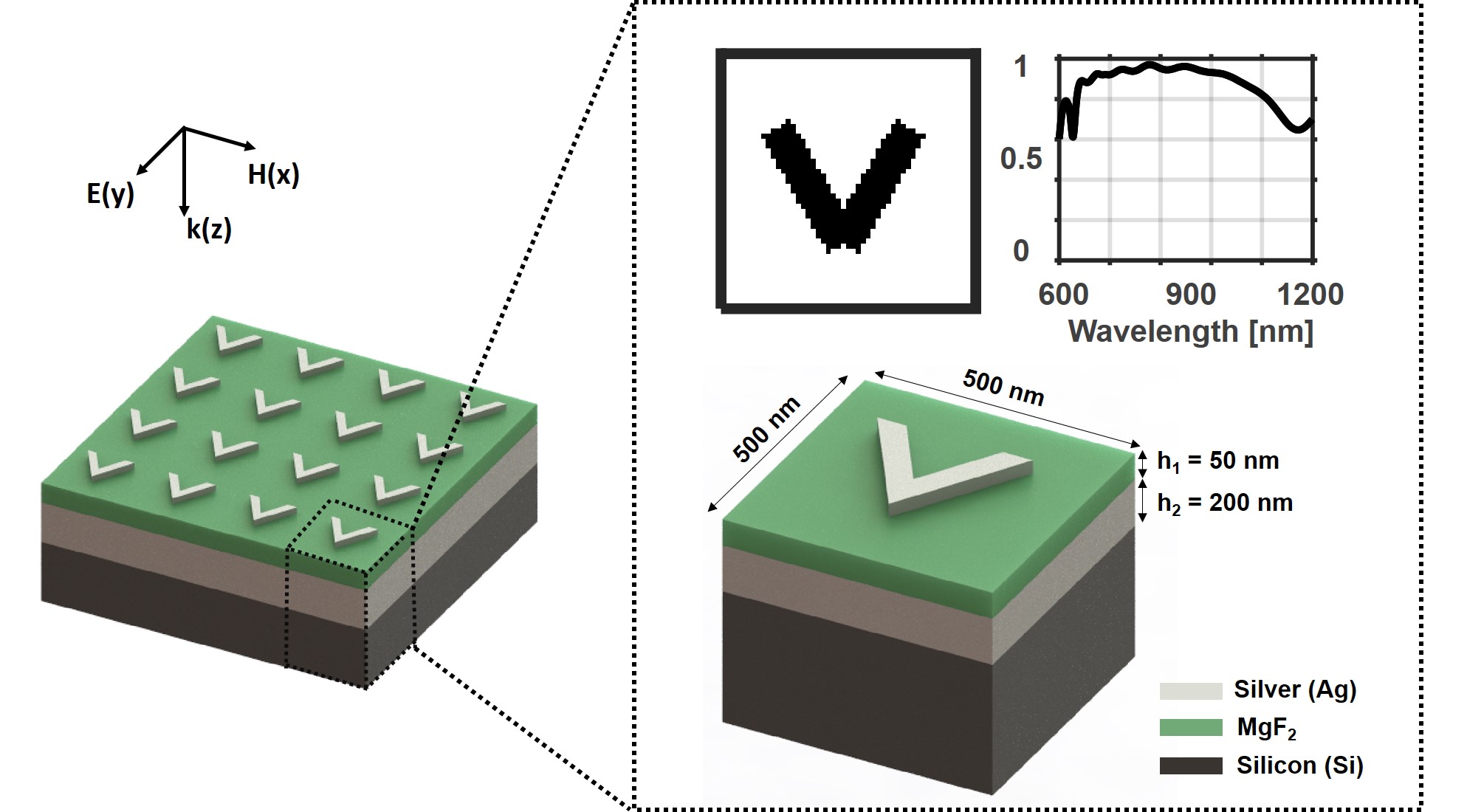}
    \caption{A schematic of data preparation for deep learning. Each entry in the data set is composed of reflection spectrum obtained from FDTD simulation and its corresponding cross-sectional structural design.}
    \label{fig:data preparation}
    \end{figure}
    
For the deep learning, we first collect a dataset consisting of 10,150 silver antennae with six representative shapes (circle, square, cross, bow-tie, H-shaped, and V-shaped). Each entry in the dataset is composed of a reflection spectrum with 200 spectral points and its corresponding cross-sectional structural design with 64 $\times$ 64 pixel image. 64 coarse meshes are used for both x- and y-direction for the simple calculation. The cross sectional structure designs are prepared in the form of images with a physical domain size of 500nm $\times$ 500nm. The antenna with 30nm thickness is placed on a 50 nm MgF$_2$ spacer, a 200nm silver reflector, and a silicon substrate (Figure \ref{fig:data preparation}). To obtain reflection spectra of each structure, a Finite-Difference Time-Domain (FDTD) electromagnetic simulation is performed in commercial program of FDTD Lumerical Solutions. The simulation is conducted over the whole spectral range from $f=250$THz to $500$THz and 200 spectral points are extracted. Periodic boundary conditions with periodicity of $500 $nm, are used along the x-direction, and y-direction respectively, and perfectly matched boundary conditions are used along the z-direction. At each simulation, y-polarized light is incident on the antenna with 0 incident angle. The current deep-learning setting solves designing structure problem in a fixed physical domain and fixed wavelength. Designing structure with different periodicity or wavelengths requires additional data collection or deep learning procedure.
    
    \begin{figure}[h!]
    \centering\includegraphics[width=10cm]{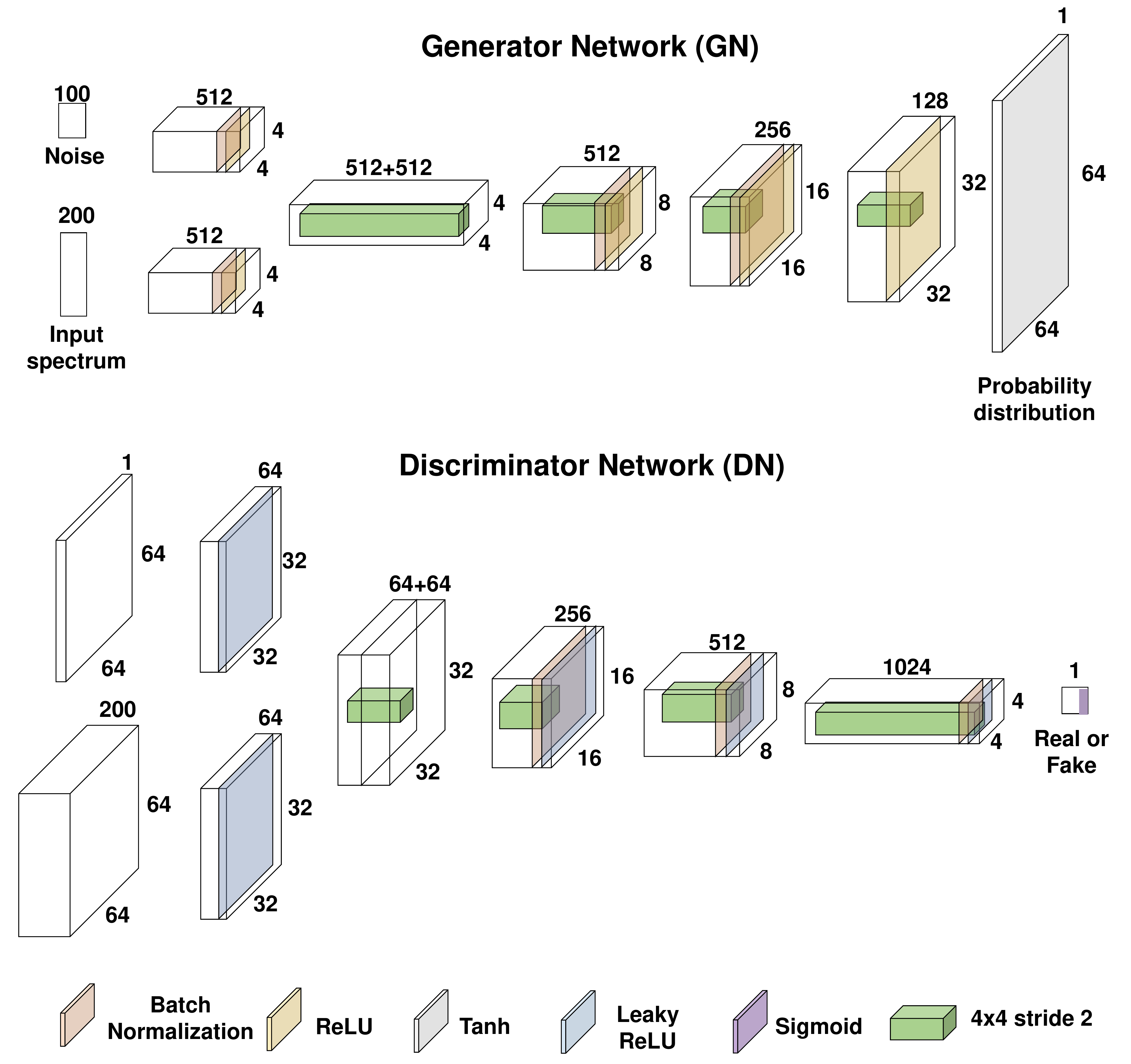}
    \caption{Schematic of the cDCGAN architecture to suggest designs of structures. GN is composed of a transposed CNN to generate structural images, and DN consists of conventional CNN to distinguish real designs from generated designs. Each layer introduces nonlinear activation functions (ReLU, Tanh, Leaky ReLU and Sigmoid) according to the guideline of Radford et al \cite{radford2015unsupervised}. }
    \label{fig:cDCGAN architecture}
    \end{figure}
    
As a next step, we implement cDCGAN algorithm using pytorch framework. The cDCGAN architecture to design nanophotonic structures is presented in Figure \ref{fig:cDCGAN architecture}. GN is composed of four transposed CNN layers that consist of 1024, 512, 256, 128, and 1 channel, respectively; DN is a CNN with four hidden layers. GN takes inputs both the 100 $\times$ 1 size random noise(z) which generates appropriate design images, and the 200 $\times$ 1 size input spectrum which directs to generate a design that satisfies the condition. GN generates a design on a 64 $\times$ 64 $-$ pixels probability distribution function (PDF) in a 500 nm $\times$ 500 nm physical domain. The generated design is again fed into DN to be discriminated from ground-truth designs. GN is trained to generate a superficial authentic design to deceive DN, and DN is trained to distinguish ground-truth designs from the design generated by G; i.e., GN and DN are simultaneously trained in the direction to minimize or maximize 
    \begin{equation}
    \begin{aligned}
     & \underset{G}{\min} \, \underset{D}{\max}V(D,G) = E_{\text{x} \leftarrow P_\text{data}(x)}[\log D(x)]+E_{z \textasciitilde P_\text{z}(z)}[\log (1-D(G(z)))],
    \end{aligned}
    \end{equation}
    where $D(x)$ represents the probability that came from real design, and $D(G(z))$ represents the probability that generated design of $G(z)$ came from generated design. 
    We modify the loss function of GN \cite{pathak2016context, isola2017image} in the cDCGAN to fit our problem to
    \begin{equation}
        l_{G}=(1-\rho) \times l_{G,design} + \rho\times l_{G,adv},
        \label{eqn:generator loss},
    \end{equation}
    where $l_{G,design}$ is design loss, $l_{G,adv}$ is adversarial loss, and $\rho$  is the ratio of adversarial loss. 
    
    \begin{equation}
        l_{G,design} = \frac{1}{n}\sum\limits_{i=1}^n (Y_{i}-\hat{Y}_{i})^2,
        \label{eqn:design loss}
    \end{equation}
    is binary cross entropy (BCE) loss between the generated design $Y_i$ and ground-truth designs $\hat{Y}_{i}$.
    
    \begin{equation}
        l_{G,adv} = \sum\limits_{n=1}^N -\log D(G(\hat{X}))
        \label{eqn:adv loss}
    \end{equation}    
    represents how well G deceives D, and is also quantitatively measured by BCE Loss.
    For DN, as adversarial loss $I_{D,adv}$, we adopt the loss function of conventional cGAN \cite{lawrence1997face}, which uses the BCE criterion as

    \begin{equation}
        l_{D} = l_{D,adv} = \sum\limits_{n=1}^N (\log D(G(\hat{X}))+\log (1-D(\hat{Z})).
        \label{eqn:discriminator loss}
    \end{equation}     
    
    We optimized $\rho$ to make GN generate high-quality realistic designs. For a sufficiently low $\rho$, a competition effect cannot be expected, whereas sufficiently high $\rho$ can cause the confusion in the learning process. Therefore, an appropriate value of $\rho = 0.5$ was chosen to maximize the ability of GN to produce convincing structural designs (See Appendix for details about deep learning procedure and network optimization).
    
    After training, cDCGAN suggests designs on a 64 $\times$ 64 $-$ pixels PDF $p(i,j)$ which represents the probability that a silver antenna exists at the location $(i,j)$. During every training step, the network is trained to optimize the weights to describe correlation between the input spectrum and the PDF. To reduce the PDF to binary structural designs, the post-processing step employes Otsu\textquotesingle s method \cite{otsu1979threshold}, which determines the binary threshold $t$ that minimizes the intra-class variance $\sigma_{\omega}^2$ of the black and white pixels as

    \begin{equation}
        \sigma_{\omega}^2 (t) = \omega_{0} (t) \sigma_0^2(t)+\omega_1 (t) \sigma_1^2(t),
        \label{eqn:ostu}
    \end{equation}       
    where $\omega_0$ and $\omega_1$ represent the weights for the probabilities of two classes separated by $t$, $\sigma_0^2$ is the variance of black pixels, and $\sigma_1^2$ is the variance of the white pixels. Therefore, for a given reflection spectrum, cDCGAN produces a PDF, which is then converted to binary designs in the post-processing step. At each step of training, 2,000 validation samples are used to validate the trained network. The average BCE Loss of validation set converged to $5.564$ $\times$ $10^{-3}$ after 1,000 epochs.

\subsection{Network evaluation}

    \begin{figure}[h!]
    \centering\includegraphics[width=10cm]{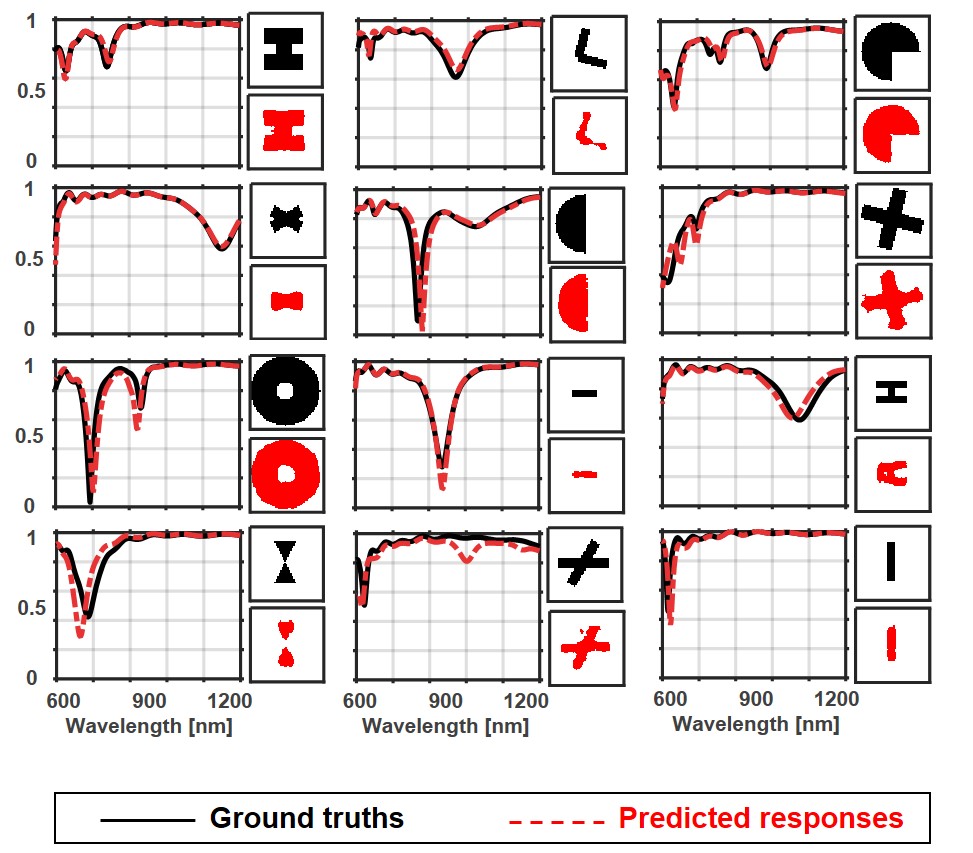}
    \caption{12 examples of cDCGAN suggested images and their simulation results. Each panel is composed of reflection spectra and their corresponding structural cross-sectional images. The upper-right structural images are ground-truth designs (black) and the lower-right images are suggested images by cDCGAN (red). The left spectra show desired input spectra (black solid lines) that we fed into the network and predicted responses obtained from the suggested designs (red dotted lines).} 
    \label{fig:cDCGAN result}
    \end{figure}

   The trained cDCGAN is evaluated on test data that were not used in previous training or validation steps. The randomly chosen test results are shown in Figure \ref{fig:cDCGAN result}. Ground-truth designs of various nanophotonic antennae (upper-right panel in Figure \ref{fig:cDCGAN result} and corresponding suggested PDFs (lower-right panel in Figure \ref{fig:cDCGAN result}) show good qualitative agreement.
    For the quantitative evaluation of the suggested PDFs, FDTD simulation based on those suggested designs are conducted. The PDFs were converted to binary designs to be imported into the simulations. Reflection spectra of the suggested images agree well with given input spectra. We introduce a mean absolute error (MAE) criterion,
    \begin{equation}
        l_{\text{error}} = \frac{1}{n}\sum_{i=1}^n |Y_i -\hat{Y}_i|
    \label{eqn:MAE}
    \end{equation}
    to quantitatively measure the accuracy of the model by comparing the predicted response with ground truth of input spectrum. The average MAE error of 12 test samples is 0.0322, which supports that the trained network can essentially provide appropriate structural design that has desired reflection spectrum.
    
\begin{figure}[h!]
    \centering\includegraphics[width=13cm]{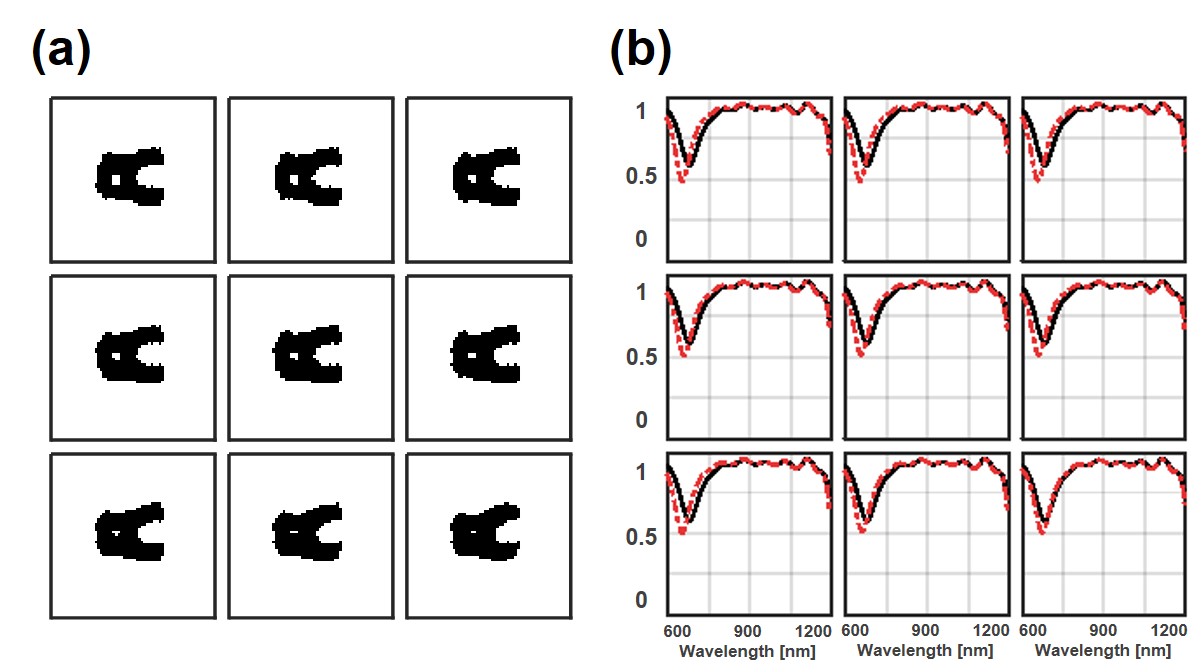}
    \caption{cDCGAN suggestion results of completely new structures of (a) triangle and (b) star shapes. The first columns are ground truths of desired designs and the second columns are PDFs suggested by the cDCGAN. The last columns show reflection spectra of input (black solid lines) and simulation result obtained from the suggested designs (red dotted lines), respectively.}
    \label{fig:latent space}
    \end{figure}

    We also explore the latent space to show that the network indeed learns the mapping between input and output. The model extract important features and such compressed representation of the features are contained in the latent space. As Radford\cite{radford2015unsupervised} et al. did, we explore the latent space by slightly changing random noise of the input \ref{fig:latent space}. Interpolation between two (first and last) random points in the latent space generates 9 different images with smooth transition. The results indicate that generator network provides smooth transition in generated images, and all of them satisfy the reflection condition well. From the Fig.~\ref{fig:latent space}, we also conclude that the reflection spectra is preserved within the manufacturing tolerance range. 
    
    \begin{figure}[h!]
    \centering\includegraphics[width=13cm]{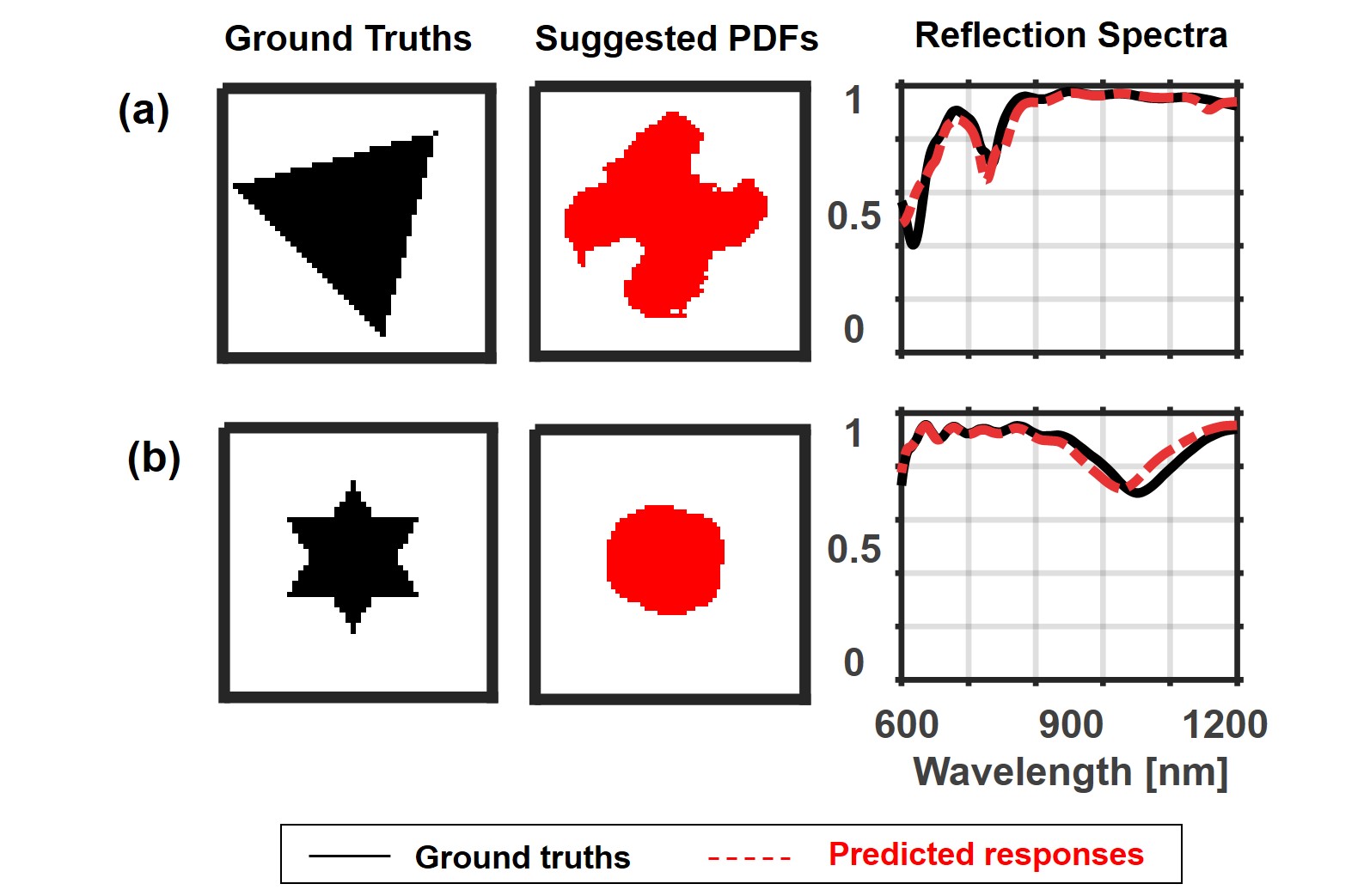}
    \caption{cDCGAN suggestion results of completely new structures of (a) triangle and (b) star shapes. The first columns are ground truths of desired designs and the second columns are PDFs suggested by the cDCGAN. The last columns show reflection spectra of input (black solid lines) and simulation result obtained from the suggested designs (red dotted lines), respectively.}
    \label{fig:cDCGAN result with new structures}
    \end{figure}
    
    We also test our cDCGAN with completely new structures of triangle and star antennae whose shapes have never been contained in the training and validation dataset (Figure \ref{fig:cDCGAN result with new structures}a ,b). The cDCGAN generated new designs that had distorted forms of the antennae that were used for training. The results imply that our cDCGAN can suggest any designs that are not constrained by structural parameters. The generated images are different from the ground-truth designs, but generate reflection spectra that are similar to input reflection spectra. This is because of the non-uniqueness of correlation between optical properties and designs: several different designs can have the same optical property. Among the several possible designs, the generated results are most likely to be found in areas that do not deviate much from the trained dataset space. 
    
    \begin{figure}[h!]
    \centering\includegraphics[width=10cm]{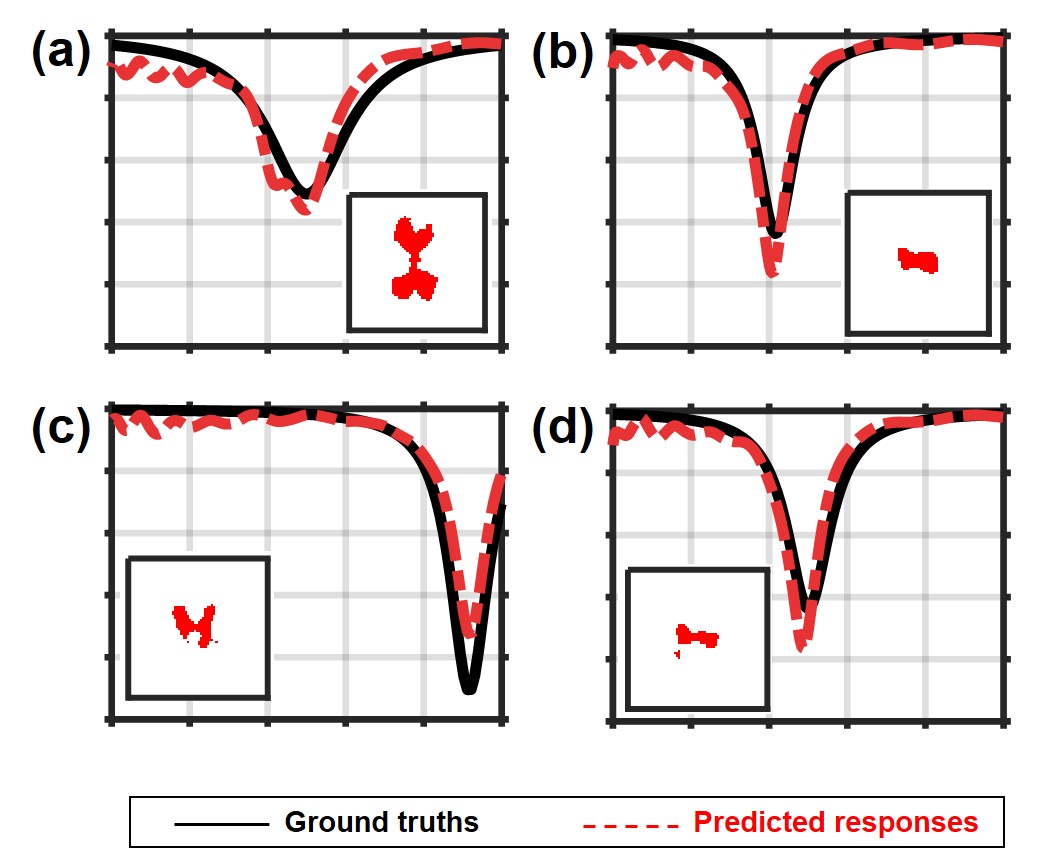}
    \caption{cDCGAN suggestion results of hand-drawn spectra with Lorentzian-like function of (a) $a=120, b=900, c=150$, (b) $a=70, b=850, c=70$, (c) $a=100,b=1150,c=70$, and (d) $a=80,b=900,c=80$ respectively. The inset in each figure shows suggested PDFs by cDCGAN. Black solid lines indicate input Lorentzian-like function that we fed to the network and red dotted lines indicate simulation results obtained from the suggested PDFs.}
    \label{fig:cDCGAN result with Lorentzian}
    \end{figure}
    
    Finally, our cDCGAN is further tested with randomly generated hand-drawn spectra of Lorentzian-like function: 
    \begin{equation}
    f(x)=\frac{2a}{\pi}\frac{c}{4(x-b)^2+c^2},
    \label{eqn:lorenzian}
    \end{equation}
    with three cases of (a) $a=120, b=900, c=150$, (b) $a=70, b=850, c=70$, (c) $a=100,b=1150,c=70$, and (d) $a=80,b=900,c=80$. For each case, the generated images and their corresponding reflection responses are shown in Figure \ref{fig:cDCGAN result with Lorentzian}. The MAE of the reflection spectrum for four examples are (a) 0.0496, (b) 0.0396, (c) 0.0409, and (d) 0.0408, respectively. The predicted responses show reasonably good agreement with input spectrum in terms of overall behaviors. Most interestingly, the generated images (insets in Figure \ref{fig:cDCGAN result with Lorentzian}a-d) are much deviated from the shapes that are used for training. Such extraordinary structural shapes are not constrained by predefined structures and even not describable; this is the key advantages of our method over previous study that can only suggest given structural parameters. The results also indicate that cDCGAN actually learns well the correlation between structural designs and their overall optical responses, and hence can be widely used to systematical design of nanophotonic structures.

\section{Conclusion}
    In conclusion, we demonstrate the first use of a cDCGAN to design nanophotonic structures. The two networks of GN and DN in cDCGAN competitively learn to suggest appropriate designs of nanophotonic structures that have desired optical properties of reflection. Our cDCGAN is not limited to suggesting predefined structures, but can also generate new designs. It has numerous design possibility with $2^{64 \times 64} = 2^{4096}$ degrees of freedom. Although our examples set the thickness of each layers and the material type of antenna, they can also be added as output parameters to be suggested. This modification would allow artificial intelligence to be used to design nanophotonic devices completely independently, and would thereby greatly reduce the time and computational cost of designing them manually. We believe that our research findings will lead to rapid development of nanophotonics by solving the main problem of designing structures.

\section*{Appendix}
\setcounter{equation}{0}
\setcounter{figure}{0}
\renewcommand{\thesubsection}{A.\arabic{subsection}}
\renewcommand{\theequation}{A.\arabic{equation}}
\renewcommand{\thefigure}{A.\arabic{figure}}
\subsection{Details on data preparation and deep learning procedure}
    A data set composed of 10,150 simulation results are prepared for deep-learning. Data on six representative structures of Circle, Square, Cross, Bow-tie, H-shaped, and V-shaped structures are collected. Among the total 10,150 original data, 80\% of the data are used for training, others 20\% are used for validation, and 20 of them are used for testing. Our conditional deep convolutional generative adversarial network (cDCGAN) produces an appropriate model that learns the correlation between design images and reflection spectrum of training dataset by adjusting the parameters of network. After learning the correlation of training dataset, the network is validated by new data that are not included training dataset to evaluate the model; i.e. validation dataset estimates how well the model is trained and thus provides the accuracy. After several training steps, the final trained network is selected to minimize the accuracy of validation data. Finally, the network is tested by randomly chosen reflection spectrum.

    We use Pytorch framework for the deep learning and adopt mini-batch gradient descent algorithm which splits the training dataset into small bathes with batch size of 64. Therefore, in every iteration, each batch containing 64 samples is used to calculate model accuracy and update model parameters. After training all the 128 batches (i.e. after one epoch), the total loss is calculated as the average of the losses in each batch. To evaluate the accuracy of the model, we adopt loss criterion of Binary Cross Entropy with Logits Loss (BCE Loss, L).

    \begin{equation}
        L=l(x,y)={l_1,...,l_N}^T, l_n = -w_n[t_n\log\sigma(x_n)+(1-t_n)\log(1-\sigma(x_n))]
    \end{equation}
    Here, N is the batch size, $t_i$, $x_i$ are the target and input to be calculated as loss. In our case, BCE Loss is appropriate to evaluate the model; this is because adversarial loss for GN and DN evaluate the authenticity of the images, and the output images provided by GN is binary images with 1 being an antenna and 0 being an air.

\subsection{Network Optimization}

    \begin{figure}[h!]
    \centering\includegraphics[width=10cm]{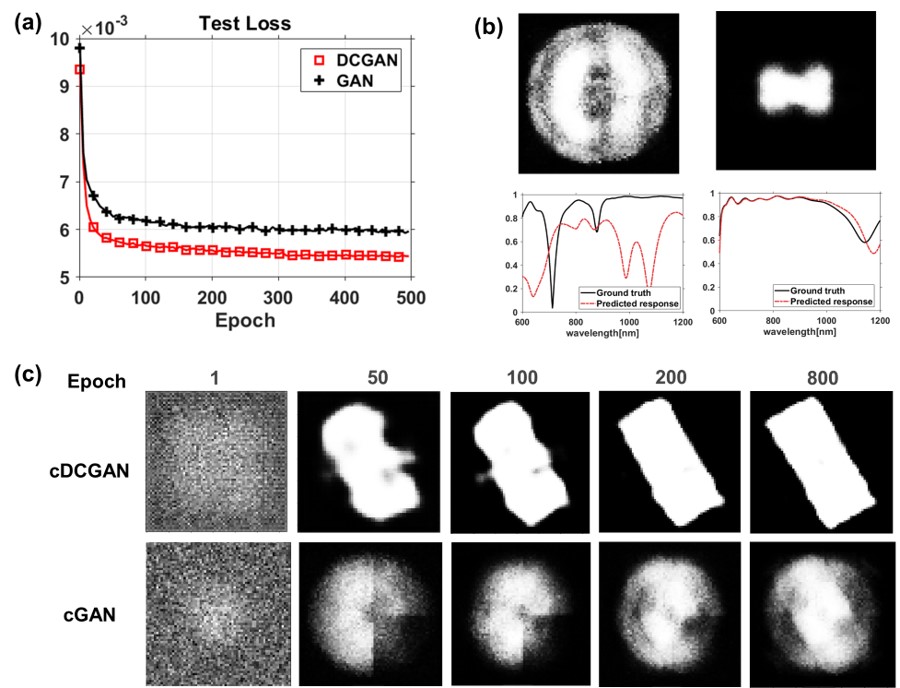}
    \caption{Comparison between cGAN and cDCGAN. (a) The average loss of 2,140 validation data of our cDCGAN (red squares) and cGAN consisting of 4 hidden layers of generator (black crosses). (b) cGAN prediction output designs and their reflection spectrum after 1,000 epochs. (c) Generated images during the training for cDCGAN (upper row) and cGAN (lower row) after several epochs. }
    \label{fig:network-optimization1}
    \end{figure}

    DCGAN is proposed to solve unstability problem of GAN, so we compared our cDCGAN with conventional generative adversarial network (cGAN) consisting of standard neural network (NN) (Figure S2). Here, the number of channels in cDCGAN is converted to the number of neuron in each layers of NN for the comparison; i.e. GN in cGAN is composed of 5 hidden layers and each layers contain 512/512, 1024, 512, 256, and 128 neurons. Figure.~\ref{fig:network-optimization1}a shows the average test loss of cDCGAN and cGAN, where best loss for to models are 5.564e-3, and 6.035e-3, respectively. After 1,000 epochs, cGAN generates noisy output designs and hence their reflection spectra do not agree well with ground truths (Figure.~\ref{fig:network-optimization1}). Figure.~\ref{fig:network-optimization1}c shows the generated images obtained after certain epochs. While cDCGAN provides realistic structural images after 200 epochs, cGAN still generates noisy images after 800 epochs. Therefore, cDCGAN efficiently learns the correlation between input spectra and their corresponding design images compared to conventional cGAN. 
    
    \begin{figure}[h!]
    \centering\includegraphics[width=10cm]{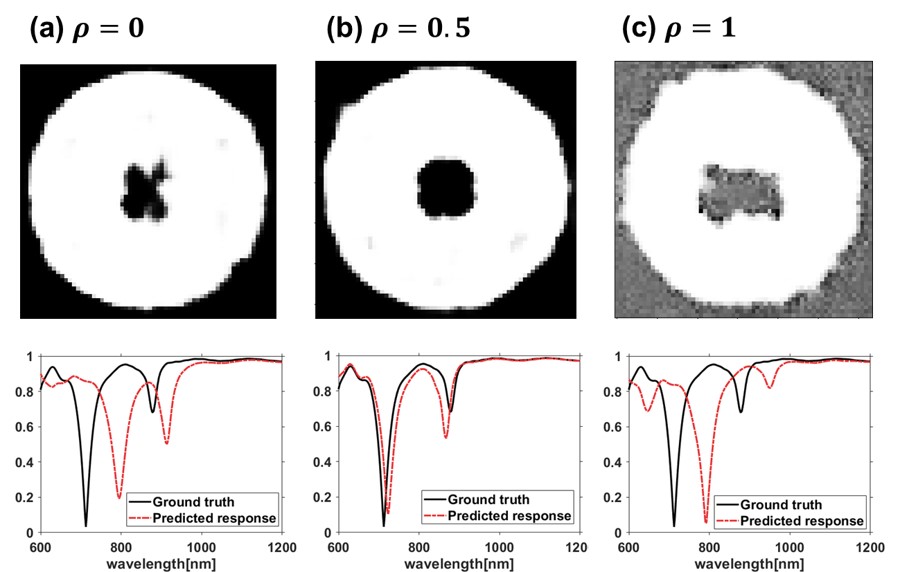}
    \caption{cDCGAN prediction results with three different value of adversarial loss ratio with (a)$\rho=0$, (b)$\rho=0.5$, and (c)$\rho=1$. First row shows suggested PDFs for circle antennas and second row shows prediction results obtained from simulations. }
    \label{fig:network-optimization2}
    \end{figure}
    
    We also optimize the weighted adversarial loss as explained in the main text. For three different ratio of $\rho=0, 0.5, 1$, networks converges to different states. The average test loss for three cases are 5.564e-3, 5.564e-3, and 1.0723e-2. Although average test loss for two cases of $\rho=0$, $\rho=0.5$ are similar, $\rho=0.5$ network can generate more authentic images compared to the cases without the adversarial loss i.e. $\rho=0$ thanks to the competition with discriminator (Figure.~\ref{fig:network-optimization2}).

\section*{Funding Information}
    This work is financially supported by the national Research Foundation grants (NRF-2017R1E1A1A03070501, NRF-2019R1A2C3003129, CAMM-2019M3A6B3030637, NRF-2018M3D1A1058998, \& NRF-2015R1A5A1037668) funded by the Ministry of Science and ICT, Korea. S.S acknowledges global Ph.D fellowship (NRF-2017H1A2A1043322) from the NRF-MSIT, Korea.

\bibliographystyle{unsrt}  
\bibliography{references}  

\end{document}